\documentclass[twocolumn]{aastex701}
\usepackage{caption}
\usepackage{xcolor}

\begin{document}

\title{Identification and photometric classification of extragalactic transients in the Vera C. Rubin Observatory's Data Preview 1}

\newcommand{\unc}{
    Department of Physics and Astronomy, 
    University of North Carolina at Chapel Hill, 
    Chapel Hill, NC 27599-3255, USA
}
\newcommand{\harvard}{Center for Astrophysics, Harvard \& Smithsonian, 60 Garden Street, Cambridge, MA 02138-1516, USA}
\newcommand{\umn}{School of Physics and Astronomy, University of Minnesota, Minneapolis, Minnesota 55455, USA}


\author[orcid=0009-0006-7990-0547]{James Freeburn}
\affiliation{\unc}
\email[show]{jfreebur@unc.edu}  

\author[0000-0002-8977-1498]{Igor Andreoni}
\affiliation{\unc}
\email{igor.andreoni@unc.edu}

\author[0000-0002-9886-2834]{Kaylee M. de Soto}
\affiliation{\harvard}
\email{kaylee.de_soto@cfa.harvard.edu}

\author[0009-0004-9687-3275]{Cristina Andrade}
\affiliation{\umn}
\email{andra104@umn.edu}


\author[0000-0002-8935-9882]{Akash Anumarlapudi} 
\affiliation{\unc}
\email{akasha@unc.edu}

\author[0000-0002-4843-345X]{Tyler Barna}
\affiliation{\umn}
\email{barna314@umn.edu}

\author[0000-0001-8544-584X]{Jonathan Carney}
\affiliation{\unc}
\email{jcarney@unc.edu}

\author[0000-0003-1314-4241]{Sushant Sharma Chaudhary}
\affiliation{\umn}
\email{ssharmac@umn.edu}

\author[0000-0002-8262-2924]{Michael W. Coughlin}
\affiliation{\umn}
\email{cough052@umn.edu}

\author[0000-0001-7129-1325]{Felipe Fontinele Nunes}
\affiliation{\umn}
\email{fonti007@umn.edu}

\author[0009-0009-5634-4244]{Sarah Teague}
\affiliation{\unc}
\email{steague1@unc.edu}

\author[0000-0002-8121-2560]{Mickael Rigault}
\affiliation{Univ Lyon, Univ Claude Bernard Lyon 1, CNRS, IP2I Lyon/IN2P3, UMR 5822, F-69622, Villeurbanne, France}
\email{m.rigault@ipnl.in2p3.fr}

\author[0000-0002-5814-4061]{V. Ashley Villar}
\affiliation{\harvard}
\affiliation{The NSF AI Institute for Artificial Intelligence and Fundamental Interactions, USA}
\email{ashleyvillar@cfa.harvard.edu}


\author[0000-0003-0042-6936]{Gloria~Fonseca~Alvarez}
\affiliation{NSF NOIRLab, 950 N.\ Cherry Ave., Tucson, AZ 85719, USA}
\email{GFonsecaAlvarez@lsst.org}

\author[0000-0003-1953-8727]{Federica~B.~Bianco}
\affiliation{Department of Physics and Astronomy, University of Delaware, Newark, DE 19716-2570, USA}
\affiliation{Joseph R.\ Biden, Jr., School of Public Policy and Administration, University of Delaware, Newark, DE 19717 USA}
\affiliation{Data Science Institute, University of Delaware, Newark, DE 19717 USA}
\email{fedhere@gmail.com}

\author[0000-0001-7387-2633]{Alexandre~Boucaud}
\affiliation{Universit\'{e} Paris Cit\'{e}, CNRS/IN2P3, APC, 4 rue Elsa Morante, F-75013 Paris, France}
\email{aboucaud@apc.in2p3.fr}

\author[0000-0003-4887-2150]{Dominique~Boutigny}
\affiliation{Universit\'{e} Savoie Mont-Blanc, CNRS/IN2P3, LAPP, 9 Chemin de Bellevue, F-74940 Annecy-le-Vieux, France}
\email{boutigny@cc.in2p3.fr}

\author{Andrew~Bradshaw}
\affiliation{SLAC National Accelerator Laboratory, 2575 Sand Hill Rd., Menlo Park, CA 94025, USA}
\affiliation{Kavli Institute for Particle Astrophysics and Cosmology, SLAC National Accelerator Laboratory, 2575 Sand Hill Rd., Menlo Park, CA 94025, USA}
\email{andrewkbradshaw@gmail.com}

\author[0000-0002-1181-1621]{Hsin-Fang~Chiang}
\affiliation{SLAC National Accelerator Laboratory, 2575 Sand Hill Rd., Menlo Park, CA 94025, USA}
\email{hfc@stanford.edu}

\author{Phil~N.~Daly}
\affiliation{Vera C.\ Rubin Observatory Project Office, 950 N.\ Cherry Ave., Tucson, AZ  85719, USA}
\email{unknown@lsst.org}

\author{Felipe~Daruich}
\affiliation{Vera C.\ Rubin Observatory, Avenida Juan Cisternas \#1500, La Serena, Chile}
\email{fdaruich@lsst.org}

\author[0009-0004-4351-5968]{Guillaume~Daubard}
\affiliation{Sorbonne Universit\'{e}, Universit\'{e} Paris Cit\'{e}, CNRS/IN2P3, LPNHE, 4 place Jussieu, F-75005 Paris, France}
\email{daubard@lpnhe.in2p3.fr}

\author[0000-0002-7790-9971]{Holger~Drass}
\affiliation{Vera C.\ Rubin Observatory, Avenida Juan Cisternas \#1500, La Serena, Chile}
\email{hdrass@lsst.org}

\author[0000-0001-7178-8868]{Laurent~Le~Guillou}
\affiliation{Sorbonne Universit\'{e}, Universit\'{e} Paris Cit\'{e}, CNRS/IN2P3, LPNHE, 4 place Jussieu, F-75005 Paris, France}
\email{llg@lpnhe.in2p3.fr}

\author[0000-0003-0800-8755]{Leanne~P.~Guy}
\affiliation{Vera C.\ Rubin Observatory, Avenida Juan Cisternas \#1500, La Serena, Chile}
\email{leanne.guy@noirlab.edu}

\author[0000-0003-3715-8138]{Patrick~Ingraham}
\affiliation{Steward Observatory, The University of Arizona, 933 N.\ Cherry Ave., Tucson, AZ 85721, USA}
\email{pingraham@arizona.edu}

\author[0000-0002-5751-3697]{M.~James~Jee}
\affiliation{Department of Astronomy, Yonsei University, 50 Yonsei-ro, Seoul 03722, Republic of Korea}
\affiliation{Physics Department, University of California, One Shields Avenue, Davis, CA 95616, USA}
\email{mkjee@yonsei.ac.kr}

\author[0000-0003-4833-9137]{Steven~M.~Kahn}
\affiliation{Physics Department,  University of California, 366 Physics North, MC 7300 Berkeley, CA 94720, USA}
\email{stevkahn@berkeley.edu}

\author[0000-0002-5261-5803]{Yijung~Kang}
\affiliation{Kavli Institute for Particle Astrophysics and Cosmology, SLAC National Accelerator Laboratory, 2575 Sand Hill Rd., Menlo Park, CA 94025, USA}
\affiliation{Vera C.\ Rubin Observatory, Avenida Juan Cisternas \#1500, La Serena, Chile}
\email{ykang@slac.stanford.edu}

\author[0000-0001-8783-6529]{Arun~Kannawadi}
\affiliation{Department of Physics, Duke University, Durham, NC 27708, USA}
\affiliation{Department of Astrophysical Sciences, Princeton University, Princeton, NJ 08544, USA}
\email{arunkannawadi@astro.princeton.edu}

\author[0000-0001-9395-4759]{Lee~S.~Kelvin}
\affiliation{Department of Astrophysical Sciences, Princeton University, Princeton, NJ 08544, USA}
\email{lkelvin@astro.princeton.edu}

\author{Didier~Laporte}
\affiliation{Sorbonne Universit\'{e}, Universit\'{e} Paris Cit\'{e}, CNRS/IN2P3, LPNHE, 4 place Jussieu, F-75005 Paris, France}
\email{didier.laporte@lpnhe.in2p3.fr}

\author{Shuang~Liang}
\affiliation{SLAC National Accelerator Laboratory, 2575 Sand Hill Rd., Menlo Park, CA 94025, USA}
\email{sliang92@stanford.edu}

\author[0000-0002-4122-9384]{Nate~B.~Lust}
\affiliation{Department of Astrophysical Sciences, Princeton University, Princeton, NJ 08544, USA}
\email{nlust@astro.princeton.edu}

\author[0000-0003-2866-3802]{Mostafa~Lutfi}
\affiliation{Vera C.\ Rubin Observatory Project Office, 950 N.\ Cherry Ave., Tucson, AZ  85719, USA}
\email{mlutfi@lsst.org}

\author[0000-0003-2384-2377]{Gabriele~Mainetti}
\affiliation{CNRS/IN2P3, CC-IN2P3, 21 avenue Pierre de Coubertin, F-69627 Villeurbanne, France}
\email{gabriele.mainetti@in2p3.fr}

\author[0000-0002-2598-0514]{Andr\'es~A.~Plazas~Malag\'on}
\affiliation{SLAC National Accelerator Laboratory, 2575 Sand Hill Rd., Menlo Park, CA 94025, USA}
\affiliation{Kavli Institute for Particle Astrophysics and Cosmology, SLAC National Accelerator Laboratory, 2575 Sand Hill Rd., Menlo Park, CA 94025, USA}
\email{plazas@stanford.edu}

\author{Felipe~Menanteau}
\affiliation{NCSA, University of Illinois at Urbana-Champaign, 1205 W.\ Clark St., Urbana, IL 61801, USA}
\email{felipe@illinois.edu}

\author{David~J.~Mills}
\affiliation{Vera C.\ Rubin Observatory Project Office, 950 N.\ Cherry Ave., Tucson, AZ  85719, USA}
\email{dmills@lsst.org}

\author{Marc~Moniez}
\affiliation{Universit\'{e} Paris-Saclay, CNRS/IN2P3, IJCLab, 15 Rue Georges Clemenceau, F-91405 Orsay, France}
\email{marc.moniez@ijclab.in2p3.fr}

\author[0000-0003-3827-4691]{Erfan~Nourbakhsh}
\affiliation{Department of Astrophysical Sciences, Princeton University, Princeton, NJ 08544, USA}
\email{erfan@princeton.edu}

\author{Russell~E.~Owen}
\affiliation{University of Washington, Dept.\ of Astronomy, Box 351580, Seattle, WA 98195, USA}
\email{rowen@uw.edu}

\author[0000-0002-4753-3387]{Maria~T.~Patterson}
\affiliation{University of Washington, Dept.\ of Astronomy, Box 351580, Seattle, WA 98195, USA}
\email{maria.t.patterson@gmail.com}

\author[0000-0001-5471-9609]{John~R.~Peterson}
\affiliation{Department of Physics and Astronomy, Purdue University, 525 Northwestern Ave., West Lafayette, IN  47907, USA}
\email{peters11@purdue.edu}

\author[0000-0002-1431-9245]{Wouter~van~Reeven}
\affiliation{Vera C.\ Rubin Observatory, Avenida Juan Cisternas \#1500, La Serena, Chile}
\email{wvanreeven@lsst.org}

\author[0000-0001-8239-3079]{Vincent~J.~Riot}
\affiliation{Lawrence Livermore National Laboratory, 7000 East Avenue, Livermore, CA 94550, USA}
\email{riot1@llnl.gov}

\author[0009-0009-2677-5537]{William~Roby}
\affiliation{Caltech/IPAC, California Institute of Technology, MS 100-22, Pasadena, CA 91125-2200, USA}
\email{roby@ipac.caltech.edu}

\author[0000-0002-9238-9521]{David~Sanmartim}
\affiliation{Vera C.\ Rubin Observatory, Avenida Juan Cisternas \#1500, La Serena, Chile}
\email{dsanmartim@lsst.org}

\author{Jacques~Sebag}
\affiliation{Vera C.\ Rubin Observatory, Avenida Juan Cisternas \#1500, La Serena, Chile}
\email{jsebag@lsst.org}

\author[0000-0003-4734-2019]{Nima~Sedaghat}
\affiliation{University of Washington, Dept.\ of Astronomy, Box 351580, Seattle, WA 98195, USA}
\email{nimaseda@uw.edu}

\author[0000-0003-4058-5202]{Richard~A.~Shaw}
\affiliation{Space Telescope Science Institute, 3700 San Martin Drive, Baltimore, MD 21218, USA}
\email{shaw@stsci.edu}

\author[0009-0000-6778-7168]{Alysha~Shugart}
\affiliation{Vera C.\ Rubin Observatory, Avenida Juan Cisternas \#1500, La Serena, Chile}
\email{alysha.shugart@noirlab.edu}

\author[0000-0002-9589-1306]{Krzysztof~Suberlak}
\affiliation{University of Washington, Dept.\ of Astronomy, Box 351580, Seattle, WA 98195, USA}
\email{suberlak@uw.edu}

\author[0000-0001-9445-1846]{John~D.~Swinbank}
\affiliation{ASTRON, Oude Hoogeveensedijk 4, 7991 PD, Dwingeloo, The Netherlands}
\affiliation{Department of Astrophysical Sciences, Princeton University, Princeton, NJ 08544, USA}
\email{swinbank@astron.nl}

\author[0000-0001-6268-1882]{Dan~S.~Taranu}
\affiliation{Department of Astrophysical Sciences, Princeton University, Princeton, NJ 08544, USA}
\email{dtaranu@princeton.edu}

\author[0000-0002-4557-6682]{Charlotte~Ward}
\affiliation{Department of Astronomy and Astrophysics, The Pennsylvania State University, 525 Davey Lab, University Park, PA 16802, USA}
\email{cvw5890@psu.edu}

\author[0000-0003-1989-4879]{Christopher~Z.~Waters}
\affiliation{Department of Astrophysical Sciences, Princeton University, Princeton, NJ 08544, USA}
\email{czw@astro.princeton.edu}

\author[0000-0001-7113-1233]{W.~M.~Wood-Vasey}
\affiliation{Department of Physics and Astronomy, University of Pittsburgh, 3941 O'Hara Street, Pittsburgh, PA 15260, USA}
\email{wmwv@pitt.edu}

\begin{abstract}

The Vera C. Rubin Observatory will soon survey the southern sky, delivering a depth and sky coverage that is unprecedented in time domain astronomy.  As part of commissioning, Data Preview 1 (DP1) has been released.  It comprises a LSSTComCam observing campaign between November and December 2024 with multi-band imaging of seven fields, covering roughly 0.4 square degrees each, providing a first glimpse into the data products that will become available once the Legacy Survey of Space and Time begins.  In this work, we search three fields for extragalactic transients.  We identify eight new likely supernovae, and three known ones from a sample of 369,644 difference image analysis objects.  Photometric classification using \texttt{Superphot+} assigns sub-classes with $>$95\% confidence to only one SN~Ia and one SN~II in this sample.  Our findings are in agreement with supernova detection rate predictions of $15\pm4$ supernovae from simulations using \texttt{simsurvey}. The supernova detection rate in the data is possibly affected by the lack of suitable templates.  Nevertheless, this work demonstrates the quality of the data products delivered in DP1 and indicates that the Rubin Observatory's Legacy Survey of Space and Time (LSST) is well placed to fulfill its discovery potential in time domain astronomy.  

\end{abstract}

\keywords{\uat{Supernovae}{1668} --- \uat{Transient detection}{1957} --- \uat{Time domain astronomy}{2109} -- \uat{Surveys}{1671}}

\section{Introduction} 

The NSF-DOE Vera C. Rubin Observatory's Legacy Survey of Space and Time \citep[LSST;][]{Ivezic2019} is set to revolutionize time-domain astronomy, exploring a new parameter space in depth and sky coverage.  This will be enabled by the largest camera in the world, LSSTCam, which will collect 3.2 gigapixels per exposure with an instantaneous field of view of 9.6 square degrees \citep{Kahn2010SPIE,10.71929/rubin/2571927}, mounted onto the 8.4-m Simonyi Survey Telescope.  Thanks to the unprecedented volumetric survey speed, the Rubin Observatory will collect a vast sample of extragalactic transients such as supernovae, tidal disruption events, active galactic nuclei, and potentially entirely new classes.  Up to 10 million alerts per night are expected to be produced by Rubin, each one indicating a change in flux compared to reference images (also referred to as `templates'). For comparison, this is more than an order of magnitude larger than the number of alerts generated nightly by the 47-square-degrees Zwicky Transient Facility \citep[ZTF;][]{2019PASP..131a8002B}.  In fact, the amount of data produced by Rubin will rapidly match all the previous astronomical images taken in human history combined.

This wealth of data, however, introduces significant challenges for data processing and analysis.  Moreover, there will not be enough resources to spectroscopically classify all the transients detected by Rubin.  Thus, photometric classification, and an understanding of transient detection efficiency, will be essential for deriving volumetric rates of various transient classes and their evolution across cosmic time.

At the time of this work, full survey operations for LSST have not yet begun.  As part of the early science program, Rubin Data Preview 1 (DP1) has been released.  Rubin DP1 provides a glimpse into real Rubin data in the form of individual exposures, difference imaging, coadded images, and catalogues obtained from the LSST Commissioning Camera \citep[LSSTComCam;][]{SITCOMTN-149}.  LSSTComCam includes nine CCDs with an instantaneous field-of-view of 0.44 square degrees.  Rubin DP1 includes observations of seven distinct fields, acquired between November and December 2024 in $ugrizy$ filters.

While limited in scope compared to the LSST main survey, Rubin DP1 provides insights into what can be expected during full survey operations.  A complementary study, \citet{2025arXiv250722156D}, search the DP1 data with a custom photometric pipeline, \texttt{SLIDE} and analyse the resultant transients' host galaxies.  In this work, we identify the first sample of extragalactic transients in Rubin data using data products from the Rubin science pipelines \citep{10.71929/rubin/2570545}, attempt photometric classification, and compare the detection rates with expectations from the literature.  

The paper is organized as follows: the dataset is described in Section\,\ref{sec:data}, and the transient search methods are presented in Section\,\ref{sec: transient search}; in Section\,\ref{sec: discussion}, we discuss the results of transient identification, photometric classification, and comparison of our analysis with expected transient detection rates. Our conclusions are presented in Section\,\ref{sec:conclusion}.

\section{Data}
\label{sec:data}

Rubin DP1 includes observations of seven distinct 0.4 square degree fields, between November and December 2024.  We refer to individual images (and their corresponding difference images) as `visits'.  Each night of observations for a given field includes multiple visits, as summarized in Table~\ref{tab:fields}.  We define an `epoch' as all visits to a specific field taken on a single night.  Included in the data products are sources (individual detections) and objects (detections grouped based on their location), extracted from the LSST science pipelines \citep{10.71929/rubin/2570545}, from both difference imaging analysis and coadded images.  In this work, we aim to identify extragalactic transients.  We therefore exclude from this analysis the Galactic fields --- 47 Tuc Globular Cluster, Low Galactic Latitude Field, Fornax, and Seagull Nebula --- due to their crowdedness and high extragalactic extinction.

\begin{figure}
    \centering
    \includegraphics[width=\columnwidth]{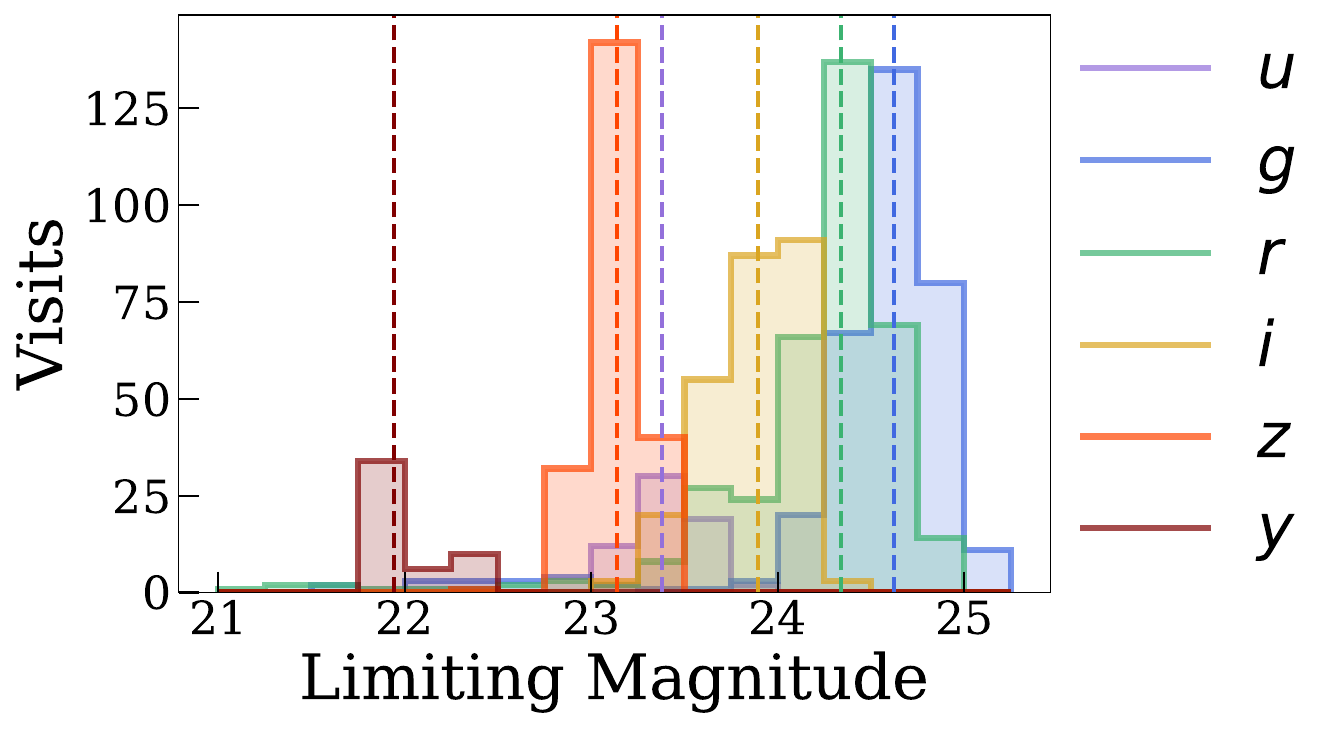}
    \caption{Distributions of the single-visit $5\sigma$ limiting magnitudes in the Rubin DP1 fields ECDFS, EDFS and LELF.  The dotted lines show the median values for each filter: $u=23.4$, $g=24.6$, $r=24.34$, $i=23.9$, $z=23.1$ and $y=21.94$.}
    \label{fig:limmag}
\end{figure}

The remaining extragalactic fields searched in this work are the Low Ecliptic Latitude Field (LELF), Extended Chandra Deep Field South (ECDFS), and Euclid Deep Field South (EDFS).  The coverage of these fields is detailed in Table~\ref{tab:fields}.  The distribution of limiting magnitudes for this dataset is shown in Figure~\ref{fig:limmag}.  

\begin{deluxetable*}{llccccrrrrrr}
\tablecaption{Rubin DP1 extragalactic fields analyzed in this work. For each field, the table presents the central coordinates, the Galactic latitude, the Galactic extinction, the number of epochs as defined in Section\,\ref{sec:data}, the mean number of visits and the single-visit median depth ($5\sigma$) per filter. \label{tab:fields}}
\tablehead{
\colhead{Field} & \colhead{Coordinates} & \colhead{Gal. Latitude} & \colhead{E(B-V)} &\colhead{Epochs} & \colhead{Mean visits} & \multicolumn{6}{c}{Median $5\sigma$ depth} \\
\colhead{} & \colhead{(J2000)} & \colhead{deg} & \colhead{} & \colhead{} & \colhead{per epoch} & \multicolumn{6}{c}{(AB mag)}\\
\multicolumn{6}{c}{} &
\colhead{$u$} & \colhead{$g$} & \colhead{$r$} & \colhead{$i$} & \colhead{$z$} & \colhead{$y$} 
}
\startdata
ECDFS & 03:32:31.2 -28:06:00.0 & -54.47 & 0.007 & 21 & 40.7 & 23.4 & 24.7 & 24.4 & 24.0 & 23.1 & 21.9 \\
EDFS  & 03:56:24.0 -48:43:48.0 & -48.48 & 0.007 & 9 & 30.2 & 23.3 & 24.5 & 24.3 & 23.9 & 23.2 & 22.2 \\
LELF  & 02:31:26.4 +06:58:48.0 & -48.18 & 0.075 & 5 & 31.8 & --   & 24.5 & 24.2 & 23.8 & 22.9 & --   \\
\enddata
\end{deluxetable*}

\section{Transient Search}
\label{sec: transient search}

\subsection{Search methodology}\label{sec:search_method}

In order to identify reliable transients, we query all difference image analysis (DIA) objects in the \texttt{DiaObject} catalog \citep{10.71929/rubin/2570319}, generated from the LSST science pipelines \citep{10.71929/rubin/2570545}.  These require at least one 5$\sigma$ detection in the DP1 difference images.  These detections are pulled irrespective of whether the image subtraction residual is positive or negative; templates for image subtraction are coadded from the third of DP1 science images with the lowest FWHM values and therefore likely include flux from long-duration transients.  A total of 369,644 DIA objects are returned from this initial query.  We describe the cuts applied to this sample in order to achieve a reasonable number of candidates for visual inspection.  The effect of these cuts are summarised in Table~\ref{tab:candidates}.

We require a minimum number of DIA detections equal to 25\% of the average number of visits per epoch for each field.  This is $\geq10$ for ECDFS and $\geq7$ for EDFS and LELF.  We require the forced photometric flux in the difference images to change by $>10\%$ in any filter, based on the \texttt{FluxLinearSlope} variable provided in the DIA data products.  A total of 3,038 candidates are returned with this query.  

To remove known active galactic nuclei (AGN) from our sample, we cross-match these 3,038 DIA objects with the Million Quasars Catalogue v8 \citep[Milliquas;][]{2023OJAp....6E..49F}, using a cross-match radius of 1''.  This removes only one source, 2021bjp, from the sample.

Solar system objects (SSOs) are a widespread contaminant in searches for extragalactic transients.  Rubin DP1 includes a catalog of identified SSOs; we remove any matches from our sample.  Additionally, we require a minimum of 30 minutes between the first and last detection for a given DIA object.  These criteria result in the removal of 73 objects, the majority of which were found in LELF due to its proximity to the ecliptic plane.

To further maximize the likelihood that the discovered transients are genuinely extragalactic, we cross-match the Rubin DIA objects with the Legacy Survey DR10 \citep[LS;][]{2019AJ....157..168D}.  Sources that are co-located (separation  $< 1''$) with cataloged sources that have a morphological classification as a point spread function (which corresponds to the \texttt{TYPE} column in LS) are discarded.  This is to remove Galactic variables and transients such as cataclysmic variables, variable stars and stellar flares as well as bright galactic nuclei.  Hostless DIA objects, without any crossmatch in LS, remain in our sample.  These cuts remove 1,896 candidates.

To remove potentially nuclear (e.g. AGN) candidates, we use \texttt{Pröst} \citep{Gagliano2025_Prost} to probabilistically associate DIA objects with galaxies from the Galaxy List for the Advanced Detector Era \citep[GLADE;][]{2018MNRAS.479.2374D} and DECam Legacy Survey \citep[DECaLS;][]{2019AJ....157..168D}.  We then use \texttt{iinuclear} \citep{iinuclear} to assess whether the objects are nuclear with respect to these galaxies.  \texttt{iinuclear} probabilistically determines whether a transient is nuclear based on the localization of the host nucleus in LS and transient's position.  This cut removed 103 objects.  While this cut may potentially remove some non-nuclear candidates, this achieves a pure sample without requiring follow-up spectroscopy.  This is necessary as, with a six-month gap between the observing campaign and DP1 data release, any transients would likely have faded beyond detectability.

Our criteria yield 965 final candidates for visual inspection.  We generate their light curves using forced PSF photometry from the \texttt{ForcedSourceOnDiaObject} catalog \citep{10.71929/rubin/2570321}.  A candidate is rejected if the detection with the highest signal-to-noise ratio (SNR) shows a bad subtraction in the difference image.  We also removed candidates if their light curve is consistent with noise from forced photometry or has sporadic detections that deviate from zero.  This is indicative of poor image subtraction.  These candidates were therefore removed to maintain the sample's purity.  A total of 11 candidates passed this visual inspection.  The \texttt{reliability} column in \texttt{DiaSource} catalogue \citep{10.71929/rubin/2570323}, a measure of a DIA source's reliability between 0 and 1 which is calculated from morphological information extracted from the source and image in addition to properties of the Telescope and Camera.  Once science operations begin, this column will likely be used to significantly reduce the required number of visual inspections to remove artefacts and bad subtractions.  However, as it is still under active development, we did not include it in our selection criteria.

\subsection{Flux offset estimation}\label{sec:flux_offsets}

\begin{figure*}[t]    
    \centering
    \includegraphics[width=\textwidth]{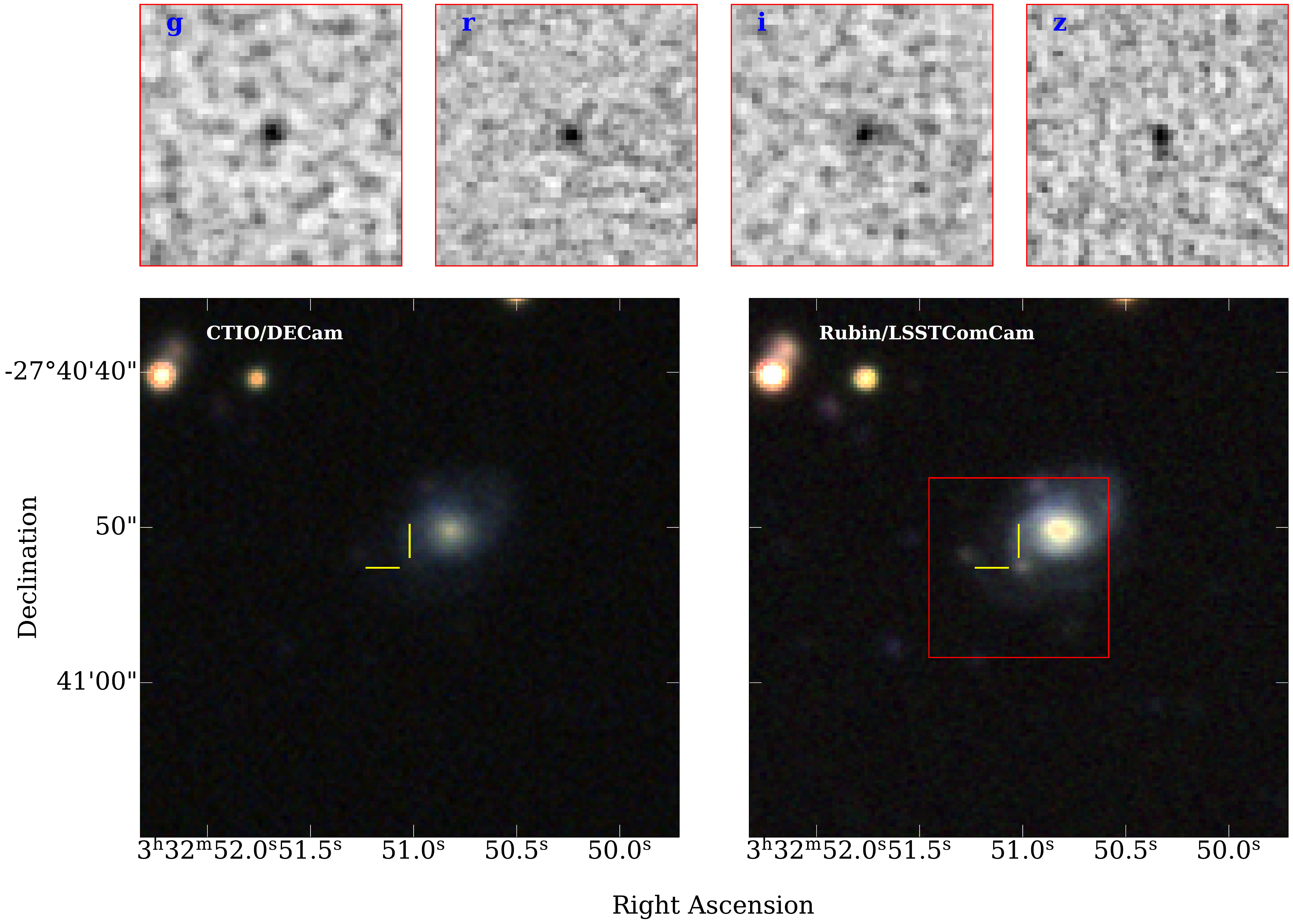}
    \caption{Coadded images of the location of 2024aigj from DECaLS (bottom left) and the Rubin DP1 templates (bottom right).  We show these images in color using $gri$ filters and indicate the location of the transient with yellow cross-hairs. The transient is present in the Rubin DP1 templates but not in the DECaLS images. \textit{Top: }Difference images of the boxed region, using the DECaLs template, isolate the transient to correct DIA object fluxes.  In these images, a dark source corresponds to a positive flux detection.}
    \label{fig:thumbnails}
\end{figure*}

DP1 template images for difference imaging are generated from a subset of observations taken during the observing campaign.  Therefore, transients identified in DP1 are often present in the templates, resulting in inaccurate flux measurements in the DIA data products.  An example of this is shown in Fig.~\ref{fig:thumbnails} with 2024aigj.  To remedy this, we perform image subtraction on, and recover forced photometry from, the DP1 templates using archival images from other instruments.  For ECDFS and EDFS, we use templates from the Dark Energy Survey DR2 \citep[DES DR2;][]{2021ApJS..255...20A}, which has coverage in analogous $griz$ filters.  For the transient in LELF, we obtain $g$ and $r$-band images from DECaLS.

We use the saccadic fast Fourier transform \citep[SFFT;][]{2022ApJ...936..157H} to perform the image subtraction and \texttt{source-extractor} \citep{1996A&AS..117..393B}, calibrated with DES DR2 for ECDFS and EDFS and Pan-STARRS1 for LELF, to extract photometry.

From this, we obtain flux offsets between the DP1 and archival DECam templates, which we use to correct photometry from the DP1 DIA data products.  This procedure is applied only to the sample of transients obtained from the search methodology described in Section~\ref{sec:search_method}.

For transients which have DP1 observations before the rise of the transient, 2024ahyy and 2024ahzc, we calculate the flux offset differently by taking the median value of the forced photometry from the first 10 days of DP1 observations for each band.  Due to the relative depth of the DP1 templates compared to the DECam templates, this allows for a more precise measurement of the flux offset. While this may affect the photometric classification results, due to only the rise phase of the light curve being sampled, we do not achieve a confident classification for these events (see Section \ref{sec:phot_class}).

\begin{deluxetable}{lrrrr}
    \tablecolumns{5}
    \caption{Summary of candidate selection criteria.\label{tab:candidates}}
    \tablehead{
    \colhead{Cut} & \multicolumn{4}{c}{Remaining candidates} \\
        & \colhead{ECDFS} & \colhead{EDFS} & \colhead{LELF} & \colhead{Total}}
    \startdata
    Initial query & 210535 & 122164 & 36945 & 369644 \\
    Selected & 1916 & 900 & 222 & 3038 \\
    Milliquas & 1916 & 900 & 221 & 3037 \\
    SSO rejection & 1906 & 895 & 163 & 2964 \\
    PSF sources & 846 & 208 & 14 & 1068 \\
    Nuclearity & 779 & 182 & 4 & 965 \\
    Inspection & 7 & 3 & 1 & 11 \\
    \enddata
\end{deluxetable}

\section{Discussion}
\label{sec: discussion}

\subsection{Identified transients}

In total, 11 candidates are obtained from the procedure described in Section~\ref{sec:search_method} and are listed in Table~\ref{tab:transients}. Three of these were previously reported on the Transient Name Server\footnote{\url{https://www.wis-tns.org/}} (TNS), with 2024aaux reported by the Zwicky Transient Facility collaboration \citep[ZTF;][]{2019PASP..131a8002B,2024TNSTR4423....1S} and 2024ahzc and 2024ahyy reported by the Young Supernova Experiment collaboration \citep[YSE;][]{2021ApJ...908..143J,2025TNSTR.975....1M}.  2024ahzc was first detected in Rubin data approximately 14 days prior to its discovery being reported on TNS.  This early detection demonstrates Rubin’s ability to identify transients early in their evolution, thanks to its combination of high cadence and sensitivity.  We expect similar examples of this to be routine in the Rubin Deep Drilling Fields, given their relatively high cadence \citep{Ivezic2019}.

Similar to this work, \citet{2025TNSAN.204....1A} provide a preliminary search of the DP1 data, identifying a number of extragalactic transients.  Of the transients reported in \citet{2025TNSAN.204....1A}, one of the three newly discovered transients, 2024aigk, is not included in this work.  Despite passing our selection criteria, it does not pass visual inspection due to poor image subtraction.  Upon further inspection of the difference images using templates from DECaLS, we identify negative flux at the location of the transient.  This indicates previous emission episodes; we therefore conclude it is a likely AGN.  Only three out of the eight previously reported transients, listed in \citet{2025TNSAN.204....1A}, are identified in this search.  The remaining five do not satisfy our selection criteria.  2024ackk and 2023yft appear in Fornax and Seagull, respectively, which are fields that are not included in this analysis.  2024ahzi and 2024ahwk failed the selection criteria due to our minimum detections requirement, which is explained in Section \ref{sec:search_method}. 2021bjp is a known AGN, present in Milliquas, and is therefore removed.

\citet{2025arXiv250722156D} identify 2024aigg, 2024ahzc, 2024ahyy and 2024aigk using a custom difference imaging and photometric pipeline, \texttt{SLIDE} which is independent of the Rubin science pipelines.  The identification of 2024aigk, which is not found in this search, potentially demonstrates the limitations of the DP1 data products.  Specifically, the presence of 2024aigk in the DP1 templates likely resulted in a lack of convincing detections in the DP1 data products.

We also note that 2024aigg was informally identified by the Rubin team in DP1 previous to its report on TNS (Taranu D. S., priv. communication, 2025).  It is also present in the ZTF alert stream as ZTF24abhdtby.  However, due to marginal detections, it was not classified as an SN in any of the ZTF alert brokers.  This demonstrates the synergy between the Rubin and ZTF alert streams which, with a combination of Rubin's depth and ZTF's cadence, will improve transient discovery and classification.

\begin{deluxetable}{lllrl}
    \tablecolumns{5}
    \tablewidth{\textwidth}
    \caption{Identified extragalactic transients in Rubin DP1. Photometric redshift measurements are obtained from the Legacy Survey DR10 photometric redshift catalog \citep{2022MNRAS.512.3662D}.}
    \tablehead{
    \colhead{IAU Name} & \colhead{Rubin ID} & \colhead{Coordinates} & \colhead{$z_{\mathrm{phot}}$} & \colhead{Reference}\\
    \colhead{} & \colhead{} & \colhead{J2000} & \colhead{} & }
    \startdata
    2024aigs & LSST-DP1-DO-591819074317582360 & 03:56:53.23 -49:06:18.06 & $0.39 \pm 0.15$ & This work\\ 
    2024aigl & LSST-DP1-DO-592913706862510093 & 03:59:24.16 -48:46:50.53 & $0.23 \pm 0.02$ & This work, $^1$\\ 
    2024aigh & LSST-DP1-DO-592915218690998602 & 03:57:17.80 -48:22:08.30 & $0.77 \pm 0.38$ & This work\\
    2024aigv & LSST-DP1-DO-609788942606139423 & 03:32:13.81 -28:28:14.40 & $0.37\pm0.06$ & This work\\
    2024ahyy & LSST-DP1-DO-609781520902651937 & 03:31:34.23 -28:24:45.45 & $0.44\pm0.10$ & $^2$\\
    2024ahzc & LSST-DP1-DO-609782208097419314 & 03:31:21.18 -28:16:47.71 & $0.06\pm0.02$ & $^2$\\
    2024aigt & LSST-DP1-DO-611253629533290657 & 03:33:41.36 -28:13:24.81 & $0.30\pm0.06$ & This work\\
    2024aigw & LSST-DP1-DO-611255210081255575 & 03:30:55.57 -27:51:58.87 & $0.32\pm0.01$ & This work\\
    2024aigg & LSST-DP1-DO-611255759837069401 & 03:32:29.94 -27:44:23.33 & $0.07\pm0.01$ & This work, $^1$\\
    2024aigj & LSST-DP1-DO-611256447031836769 & 03:32:51.02 -27:40:52.60 & $0.25\pm0.05$ & This work\\
    2024aaux & LSST-DP1-DO-648374722634973207 & 02:34:22.75 +07:12:52.67 & $0.09\pm0.01$ & $^3$\\
    \enddata
    \tablecomments{$^1$\citet{2025TNSAN.204....1A}, $^2$\citet{2025TNSTR.975....1M}, $^3$\citet{2024TNSTR4423....1S}}
    \label{tab:transients}
\end{deluxetable}

\begin{figure*}   
    \centering
    \includegraphics[width=\textwidth]{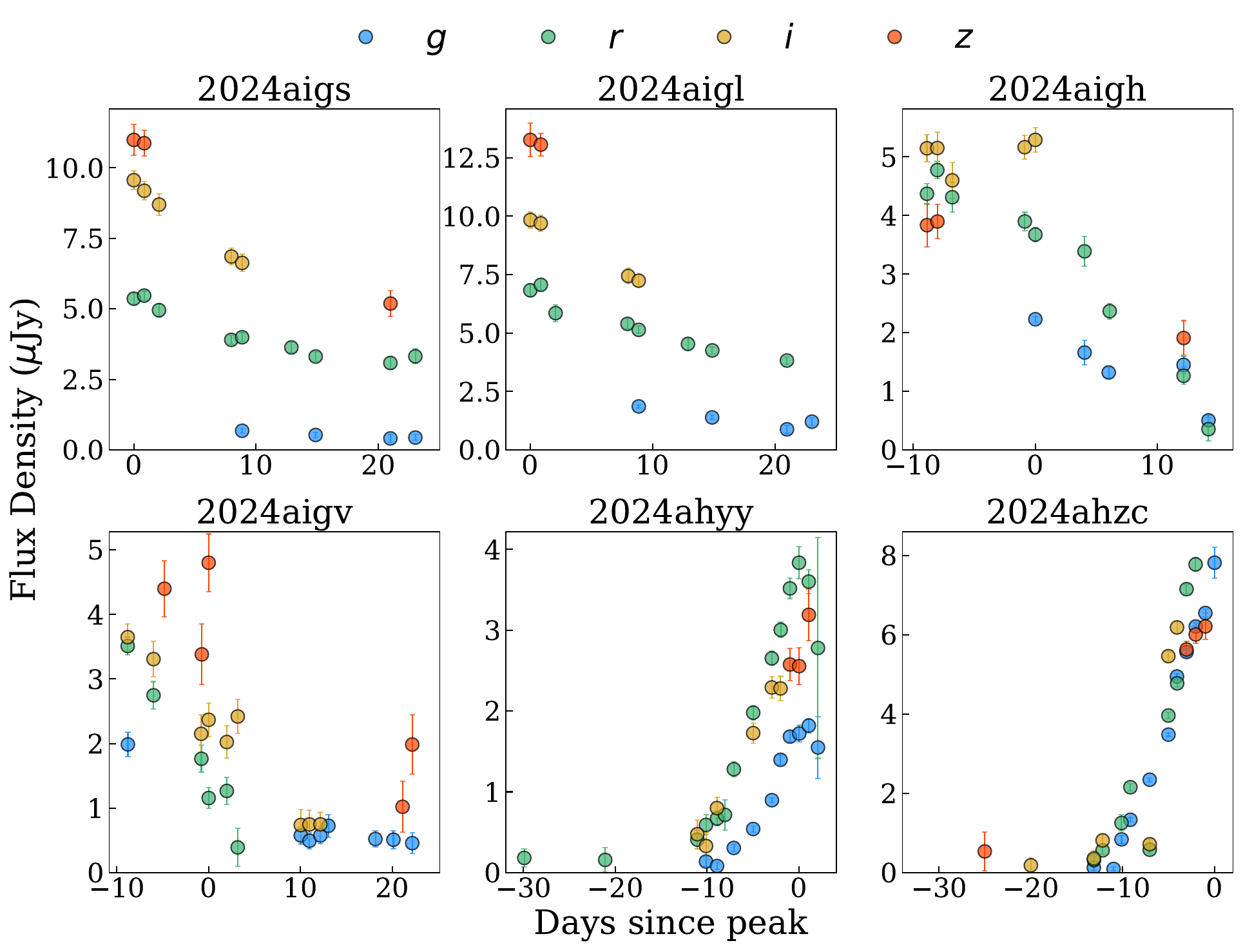}
    \caption{Multi-band light curves of extragalactic transients identified in this search. Each data point is the result of binning all DIA forced PSF photometry from the \texttt{ForcedSourceOnDiaObject} catalog \citep{10.71929/rubin/2570321} from all the visits over a single epoch. A flux offset, obtained using the method described in Section \ref{sec:flux_offsets}, has been applied to each light curve.}
    \label{fig:lcs1}
\end{figure*}

\begin{figure*}  
    \ContinuedFloat
    \centering
    \includegraphics[width=\textwidth]{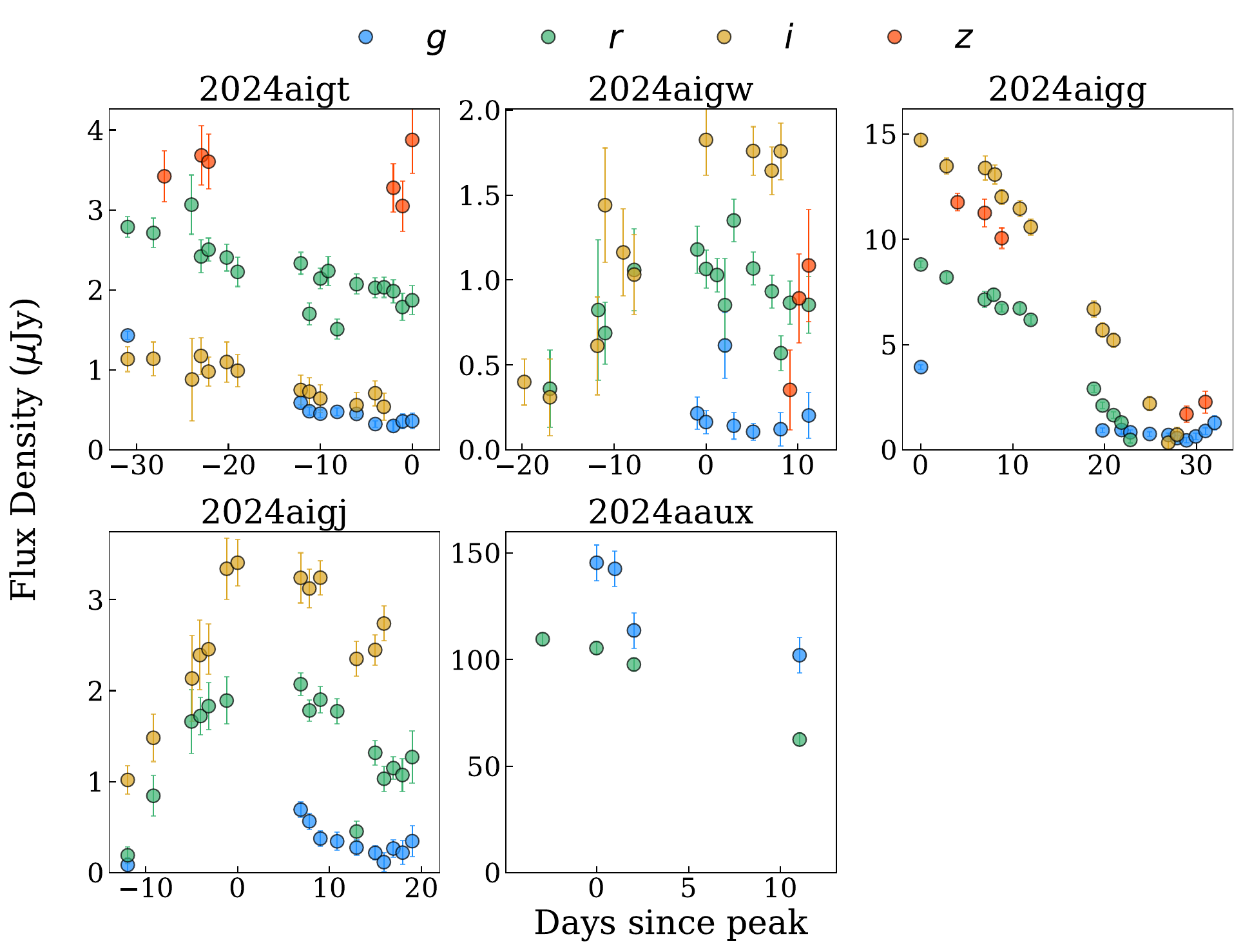}
    \caption{(continued)}
    \label{fig:lcs2}
\end{figure*}

\subsection{Photometric classification of transients}\label{sec:phot_class}

Because we eliminate nuclear, Galactic and strangely-evolving transients from the sample in Sec.~\ref{sec:search_method}, we claim that the 11 remaining transients are supernovae (SNe).  This is a reasonable assumption, as SNe are the most common class of extragalactic transients (see Section~\ref{sec:vol_rates}).  We aim to sub-classify these objects in order to determine if our observed rates are consistent with estimated volumetric rates of each SN subclass.  We lack spectra for all objects in the sample, so we instead rely on photometric classification from the subtracted multi-band light curves in Fig.~\ref{fig:lcs1}.  We use \texttt{Superphot+}~\citep{superphotplus} to fit each transient's photometry to a general SN model, and from the best-fit parameters output one of five SN subtypes: Ia, II, IIn, Ibc, or SLSN-I.  \texttt{Superphot+} excels at extracting information from sparse or partial light curves without overfitting. Because we have photometric redshift estimates for all candidates in our sample, we correct for time dilation.  We also use the pre-trained early- and full phase \texttt{Superphot+} classifiers that include peak $r$-band absolute magnitude, calculated from the modeled light curves, as an input feature.

This classifier is trained on ZTF $g$- and $r$-band photometry from a large sample of spectroscopically confirmed SNe, so while we use all {LSST}ComCam bands to jointly constrain fit parameters, we only use the parameters from two bands of data for classification. The bands are chosen such that their effective wavelengths are closest to where the emitted $g$- and $r$-band effective wavelengths are redshifted. For example, 2024ahyy has a mean photometric redshift of 0.39, which would shift the $g$- and $r$-band effective wavelengths to 694.8 and 895.9 nm, respectively. This most closely remaps to $i$- and $z$-band, so we treat the $i$- and $z$-band fit parameters as the $g$- and $r$-band inputs when using the classifier. This serves as an alternative to directly applying $K$-corrections, since we don't know the candidates' SEDs \textit{a priori}. The ``alignment'' score in Table~\ref{tab:photclass} indicates how precise the filter shift is for each object; values near one indicate a near perfect filter shift, while values closer to 0.5 indicate that the shifted wavelengths land between two filters, possibly degrading classification performance.

The parametric fitting model \citep{2019ApJ...884...83V} accounts for a rise in brightness, followed by an optional plateau (i.e. for SNe~II) and subsequent decline. For partial light curves, we cannot constrain the parameters defining regions with no observations. Therefore, we apply separate full-, early-, and late-phase classifiers to the light curves depending on phase coverage, which are trained with subsets of fit parameters. Both the full- and early-phase classifiers are reused from \cite{superphotplus}, while a new late-phase classifier was created for this analysis. When the rise is missing, we only use late-phase features, and if the decline is missing, we only use early-phase features. Only 2024aigj is classified with both early- and late-phase fit parameters.

The classification results are shown in Table~\ref{tab:photclass}. We label each object as the SN class with highest pseudo-probability, though we note that these outputs are not calibrated. SN Ia and SN II predictions tend to be underconfident, while minority class predictions tend to be overconfident (see \citealt{superphotplus} for more information). While calibrating a multi-class classifier with an imbalanced dataset is nontrivial \citep{silva_filho2023classifier}, we define a confidence metric by calibrating each class to a one-vs-rest logistic curve. We then map each object's predicted class to a confidence value using these calibration curves. We also obtain an anomaly score for each object's photometry from an isolation forest trained on the ZTF dataset's fit parameter set. Negative anomaly scores indicate potentially anomalous light curves relative to the ZTF dataset.

Two out of the eleven transients in our sample are classified as SN~Ia, three are classified as SN~II, one is classified as SN~Ibc, one is classified as SLSN-I, and four are classified as SN~IIn. This is inconsistent with the expected class fractions in \cite{2020ApJ...904...35P}, with an exceptionally high number of SNe~IIn. Two classifications are confident beyond the 95\% level and five have confidence levels $>50$\%, all predicted SNe Ia or SNe II. The predicted SNe IIn, SLSN-I, and SN Ibc all have low confidences, and all light curves except one are considered anomalies relative to the ZTF dataset. For the purposes of rate calculations, we therefore treat the transients in our sample as likely SNe rather than assigning them to any particular class.

These dubious fits and classifications could be attributed to inexact filter shifts, dust extinction from host galaxies, imperfect template subtraction, or baked-in biases between ZTF and LSSTComCam photometry.  LSST will operate at a lower observational cadence than DP1, yielding more sparsely sampled light curves. On the other hand, LSST will revisit the same fields for the full duration of a typical supernova's evolution, so we will be far less reliant on a late-phase classifier past the first month of observation.  We expect LSST's camera to have lower thermal noise and better systematics than LSSTComCam, reducing flux uncertainties. It is therefore unclear whether classification of LSST light curves will be better or worse compared to LSSTComCam.

\begin{deluxetable}{lrrrrrrrr}
    \tablecolumns{9}
    \tablewidth{\columnwidth}
    \caption{Probabilities from \texttt{Superphot+} and final classification for each transient \label{tab:photclass}.  $M_{r,\mathrm{peak}}$ is calculated from the best-fit model's peak apparent magnitude and the mean photometric redshift.  The errors associated with $M_{r,\mathrm{peak}}$ are calculated purely from the photometric redshift uncertainty.}
    \tablehead{
        \colhead{IAU Name} &
        \colhead{$M_{r,\mathrm{peak}}$} &
        \colhead{Model} &
        \colhead{Bands} &
        \colhead{Alignment} &
        \colhead{Pred.\ Class} &
        \colhead{$p_\mathrm{pred}$} &
        \colhead{Confidence} &
        \colhead{Anom. Score}
    }
    \startdata
2024aigs & $-20.00_{-0.72}^{+1.08}$ & late  & $z$ / $r$ & 0.798 & SN IIn  & 0.395 & 0.132 & -0.041 \\
\textbf{2024aigl} & \textbf{$-19.03_{-0.20}^{+0.22}$} & \textbf{late}  & \textbf{$i$ / $r$} & \textbf{0.850} & \textbf{SN Ia}   & \textbf{0.410} & \textbf{0.814} & \textbf{-0.089} \\
2024aigh & $-20.76_{-0.87}^{+1.54}$ & late  & $z$ / $i$ & 0.434 & SLSN-I  & 0.442 & 0.056 & -0.231 \\
\textbf{2024aigv} & \textbf{$-19.01_{-0.31}^{+0.36}$} & \textbf{late}  & \textbf{$z$ / $r$} & \textbf{0.780} & \textbf{SN Ia}   & \textbf{0.707} & \textbf{0.984} & \textbf{-0.072} \\
2024ahyy & $-19.13_{-0.48}^{+0.61}$ & early & $z$ / $i$ & 0.643 & SN IIn  & 0.442 & 0.133 & -0.200 \\
2024ahzc & $-15.56_{-0.58}^{+0.78}$ & early & $r$ / $g$ & 0.755 & SN IIn  & 0.339 & 0.059 & 0.005 \\
\textbf{2024aigt} & \textbf{$-18.19_{-0.37}^{+0.44}$} & \textbf{late}  & \textbf{$i$ / $r$} & \textbf{0.744} & \textbf{SN II}   & \textbf{0.503} & \textbf{0.726} & \textbf{-0.195} \\
\textbf{2024aigw} & \textbf{$-18.12_{-0.07}^{+0.08}$} & \textbf{early} & \textbf{$z$ / $r$} & \textbf{0.736} & \textbf{SN II}   & \textbf{0.712} & \textbf{0.954} & \textbf{-0.191} \\
\textbf{2024aigg} & \textbf{$-16.27_{-0.43}^{+0.54}$} & \textbf{late}  & \textbf{$r$ / $g$} & \textbf{0.715} & \textbf{SN II}   & \textbf{0.464} & \textbf{0.647} & \textbf{-0.216} \\
2024aigj & $-17.74_{-0.39}^{+0.47}$ & full  & $i$ / $r$ & 0.830 & SN Ibc  & 0.567 & 0.172 & -0.089 \\
2024aaux & $-19.12_{-0.23}^{+0.26}$ & late  & $r$ / $g$ & 0.633 & SN IIn  & 0.295 & 0.058 & -0.007 \\
    \enddata
\end{deluxetable}

\subsection{Extragalactic transient rate prediction for the main survey}\label{sec:vol_rates}

Volumetric rates of supernova subclasses can be estimated from systematic surveys.  SN~Ia and core-collapse supernovae (CCSNe) dominate survey transient detection rates due to their luminosity and large volumetric rates of $2.35\pm0.24\times10^4$\,Gpc$^{-3}$yr$^{-1}$ and $10.1^{+10}_{-3.5}\times10^4$\,Gpc$^{-3}$yr$^{-1}$, respectively \citep{2020ApJ...904...35P}.  Tidal disruption events, kilonovae, and luminous fast blue optical transients account for $<1.5\%$ of the transient detection rate \citep{2020ApJ...904...35P}, which we treat as negligible in our estimates.

To compare the number of transients we identified in DP1 with realistic expectations for supernova detection, we employ \texttt{skysurvey}\footnote{\url{skysurvey.readthedocs.io}}, a transient simulation package.  It accounts for time dilation, k-corrections, cosmic evolution and Galactic and host galaxy extinction and was previously used to probe the ZTF SN~Ia DR2 selection function (\citealt{2025A&A...694A...1R}; see \citealt{2025A&A...694A...3A} for a full description of the software). 

Using rates from \citealt{2020ApJ...904...35P}, we generate a series of 100 volume-limited samples ($z<1$) of SNe Ia, assuming a SALT2-extended light curve template \citep{2010A&A...523A...7G} as made available by \texttt{sncosmo} \citep{2016ascl.soft11017B}. The 100 iterations of the sample allow for a robust estimation of the error associated with the predicted rates.  SALT2 stretch and color parameters have been drawn assuming the \cite{2025A&A...694A...4G} and \cite{2025A&A...695A.140G} distributions. Realistic light curves are then generated according to the observational conditions present during DP1 observations.

To emulate the criteria used for DP1 transients, we combine photometry taken in the same night for each band, keeping individual detections if they exceed an SNR of 10. Based on the requirement on minimum epochs in the transient search, we require light curves to have at least 7 different observations to be marked as detected. Following these criteria, we predict $12 \pm 3$ SNe Ia.  We calculate the CCSN rate in the DP1 data of $3\pm2$ by scaling our predicted SNe Ia rate using the relative number of detections between SN~Ia and SN~II in the magnitude-limited sample reported in \citet{2020ApJ...904...35P}.  This is an approximation for Rubin data, the relative rate in \citet{2020ApJ...904...35P} calculated for the low-$z$ universe and for only $g$ and $r$ filters.  The DP1 data, in contrast, includes more filters and is probing higher redshifts.  A more detailed simulation of CCSNe would be required to estimate rates and absolute magnitude redshift evolution, given their complexity (see e.g. \cite{Vincenzi_2021}).  This corresponds to rate of $15\pm4$ all classes in the DP1 data which is consistent with the observed likely supernovae reported in Table \ref{tab:transients}.  The photometric classifications in Table~\ref{tab:photclass} are not confident enough to compare the relative rates of sub-classes.

\section{Conclusion}
\label{sec:conclusion}

The Vera C. Rubin Observatory's Legacy Survey of Space and Time will discover extragalactic transients at an unprecedented rate.  Using Rubin's data preview 1, we provide a first glimpse into this new era of time domain astronomy by conducting a search for extragalactic transients. 

After removing likely active galactic nuclei, solar system objects, and Galactic variables and transients from the sample, we identify eleven extragalactic transients, which are likely supernovae (SNe). Eight of these are newly discovered transients, and three were previously identified by other surveys and reported to TNS.  The sensitivity of this search was limited by the DP1 templates being comprised of science images, resulting in transient flux being present in these templates.

Using \texttt{Superphot+}, we attempt photometric classification of these likely SNe.  We do not confidently ($>95\%$) assign their sub-classes for all the identified SNe except for one SN~Ia and one SN~II.  With the likely decrease in observational cadence in the main survey compared to DP1, there may be further challenges in photometric classification once science operations begin.  Conversely, LSST will revisit the same fields for the full duration of a typical supernova’s evolution and LSSTCam may have reduced systematics and flux uncertainties compared to LSSTComCam which may improve photometric classification performance.  Looking forward, we will be attempting classification with frameworks that combine images and photometry \citep{junell2025applyingmultimodallearningclassify}.  Nevertheless, the results of this search are consistent with the expected SN rate of $15\pm4$ predicted with \texttt{skysurvey}.

This work demonstrates that the DP1 data products, with relatively simple selection criteria, can be sifted for extragalactic transients.  This reflects the high quality of the LSST science pipelines \citep{10.71929/rubin/2570545} which are undergoing further optimization. This search was limited by templates being constructed from images taken during the observing campaign itself.  However, with the LSST main survey, this will cease to be an issue once templates are produced in the early months of the survey. Looking forward to LSST science operations, we conclude that the Vera C. Rubin Observatory is well placed to achieve its time domain science goals.

\begin{acknowledgements}

We thank Lei Hu for helpful advice regarding image subtraction.

We also thank the anonymous reviewer for providing comprehensive feedback which has significantly improved the paper.

The Andreoni Transient Astronomy Lab is supported by the National Science Foundation award AST 2505775, NASA grant 24-ADAP24-0159, and the Discovery Alliance Catalyst Fellowship Mentors award 2025-62192-CM-19. M.W.C., S.S.C., C.A. and F.F.N acknowledge support from the National Science Foundation with grant numbers PHY-2117997, PHY-2308862 and PHY-2409481.  V.A.V. and K.d.S. acknowledge support through the David and Lucile Packard Foundation, the Research Corporation for Scientific Advancement (through a Cottrell Fellowship), the National Science Foundation under AST-2433718, AST-2407922 and AST-2406110, as well as an Aramont Fellowship for Emerging Science Research.  This work is supported by the National Science Foundation under Cooperative Agreement PHY-2019786 (the NSF AI Institute for Artificial Intelligence and Fundamental Interactions). K.d.S. thanks the LSST-DA Data Science Fellowship Program, which is funded by LSST-DA, the Brinson Foundation, the WoodNext Foundation, and the Research Corporation for Science Advancement Foundation; her participation in the program has benefited this work.

This material is based upon work supported in part by the National Science Foundation through Cooperative Agreements AST-1258333 and AST-2241526 and Cooperative Support Agreements AST-1202910 and 2211468 managed by the Association of Universities for Research in Astronomy (AURA), and the Department of Energy under Contract No. DE-AC02-76SF00515 with the SLAC National Accelerator Laboratory managed by Stanford University. Additional Rubin Observatory funding comes from private donations, grants to universities, and in-kind support from LSST-DA Institutional Members.

\end{acknowledgements}

\bibliography{ref}{}
\bibliographystyle{aasjournalv7}





\end{document}